\documentclass[a4paper,11pt]{article}
\usepackage{amssymb,graphicx,amsmath}

\makeatletter 
\renewcommand\section{\@startsection{section}{1}{\z@}              {-3.25ex\@plus -1ex \@minus -.2ex}                                    {1.5ex \@plus .2ex}                                    {\reset@font\Large\bfseries}}
\renewcommand\subsection{\@startsection{subsection}{2}{\z@}                                    {3.25ex \@plus 1ex \@minus.2ex}                                    {-1em}                                    {\reset@font\large\bfseries}}
\@addtoreset{equation}{section} \makeatother
\renewcommand{\theequation}{\thesection.\arabic{equation}}
\input{tcilatex}
\begin{document}

\setcounter{page}{0} \topmargin0pt \oddsidemargin0mm \renewcommand{%
\thefootnote}{\fnsymbol{footnote}} \newpage \setcounter{page}{0}
\begin{titlepage}
\begin{flushright}
CMS 04/06  \\
\end{flushright}
\vspace{0.2cm}
\begin{center}
{\Large {\bf Boundary form factors of the sinh-Gordon model with
Dirichlet boundary conditions at the self-dual point}}

\vspace{0.8cm} {\large  \text{Olalla
A.~Castro-Alvaredo$^{\bullet}$}}

\vspace{0.2cm}
{Centre for Mathematical Science, City University London, \\
Northampton Square, London EC1V 0HB, UK}
\end{center}
\vspace{0.5cm}

\renewcommand{\thefootnote}{\arabic{footnote}}
\setcounter{footnote}{0}

\begin{abstract}
\normalsize \noindent In this manuscript we present a detailed
investigation of the form factors of boundary fields of the
sinh-Gordon model with a particular type of Dirichlet boundary
condition, corresponding to zero value of the sinh-Gordon field at
the boundary, at the self-dual point. We follow for this the
boundary form factor program recently proposed by Z.~Bajnok,
L.~Palla and G.~Takacs \cite{BPT}, extending the analysis of the
boundary sinh-Gordon model initiated there. The main result of the
paper is a conjecture for the structure of all $n$-particle form
factors of the boundary operators $\partial_{x} \phi$ and
$(\partial_{x}\phi)^{2}$ in terms of elementary symmetric
polynomials in certain functions of the rapidity variables. In
addition, form factors of boundary ``descendant" fields have been
constructed.
\medskip
\medskip


\end{abstract}

\vfill{ \hspace*{-9mm}
\begin{tabular}{l}
\rule{6cm}{0.05 mm}\\
\text{$\bullet$ o.castro-alvaredo@city.ac.uk}\\
\end{tabular}}
\end{titlepage}
\newpage
\section{Introduction}
In the context of 1+1-integrable quantum field theories (QFTs)
form factors are defined as tensor valued functions, representing
matrix elements of some local operator $\mathcal{O}({x})$ located
at the origin between a multi-particle {\em{in}}-state and the
vacuum
\begin{equation}
F_{n}^{\mathcal{O}|\mu _{1}\ldots \mu _{n}}(\theta _{1},\ldots
,\theta _{n}):=\left\langle 0|\mathcal{O}(0)Z_{\mu _{1}}(\theta
_{1})Z_{\mu _{2}}(\theta _{2})\ldots Z_{\mu _{n}}(\theta _{n})
|0\right\rangle _{\text{in}}\,\,. \label{ff}
\end{equation}
Here $|0\rangle$ represents the vacuum state and $Z_{\mu
_{i}}(\theta _{i})$ are vertex operators which provide a
generalization of Bosonic or Fermionic algebras and allow for the
definition of a space of physical states. They carry indices
$\mu_i$, which are quantum numbers characterizing the various
particle species, and depend on the parameters $\theta_i$, which
are called rapidities. In 1+1 dimensions the energy and momentum
of the particles can be expressed in terms of the rapidity
variables as
\begin{equation}\label{em}
    E_i=m_i \cosh \theta_i \qquad \text{and} \qquad P_i=m_i \sinh \theta_i,
\end{equation}
where $m_i$ is the mass of the particle considered. The braiding
relations amongst the operators $Z_{\mu _{i}}(\theta _{i})$ and
their Hermitian conjugates are known as Faddeev-Zamolodchikov
algebra \cite{cFad,za}. For theories without back-scattering
(diagonal theories) they are given by
\begin{eqnarray}
Z_{\mu_i}(\theta _{i})Z_{\mu_{j}}(\theta _{j})&=& S_{\mu_i
\mu_j}(\theta _{ij})Z_{\mu_j}(\theta _{j})Z_{\mu_i}(\theta _{i}),
\label{Zalg1}\\
Z_{\mu_i}^{\dagger}(\theta _{i})Z_{\mu_j}^{\dagger}(\theta
_{j})&=& S_{\mu_i \mu_j}(\theta _{ij})Z_{\mu_j}^{\dagger}(\theta
_{j})
Z_{\mu_i}^\dagger(\theta _{i}),\label{Zalg2}\\
Z_{\mu_i}(\theta _{i})Z_{\mu_j}^{\dagger}(\theta _{j})&=& S_{\mu_i
\mu_j}(-\theta _{ij})Z_{\mu_j}^\dagger(\theta _{j})
Z_{\mu_i}(\theta _{i})+ 2\pi \delta_{\mu_i \mu_j}\delta(\theta
_{ij}), \label{Zalg3}
\end{eqnarray}
where $S_{\mu_i \mu_j}(\theta _{ij})$ is the 2-particle scattering
matrix and we introduced the variable
$\theta_{ij}=\theta_i-\theta_j$.

The form factor program for integrable models was pioneered by
P.~Weisz and M.~Karowski \cite{Weisz,KW} and thereafter developed
to a large extent by F.~A.~Smirnov, who also formulated some of
the consistency equations for form factors \cite{smirnovbook}. In
these seminal works, the fundamental properties of form factors in
1+1-dimensional theories were established (see
\cite{Essler:2004ht} for a recent review). It was found that the
form factors of local operators can be obtained as the solutions
to a set of consistency equations whose origin is based on
physically-motivated requirements. The solution to these equations
allows in principle the computation of all $n$-particle form
factors associated to any local field of the massive QFT.
Following this program, form factors of a large class of
integrable models have been computed (see e.g.
\cite{Weisz}-\cite{Delfino:2006te} and references therein). Once
the form factors of a certain operator are known they can be used
for many interesting applications like the computation of
correlation functions (see e.g. \!\cite{Yzam},
\cite{Delfino:1993ha}-\cite{Acerbi:1996kh} and \cite{HSG1,HSG2}),
the identification of the operator content of the perturbed
conformal field theory (CFT) (see e.g. \cite{KK,Smirnov42,DSC} and
\cite{HSG1}-\cite{Grinza:2004hr}) and the explicit computation of
quantities which characterize the underlying CFT (see e.g.
\cite{Cardy,Mussardo:1993ut,Delfino:1994ea,DSC} and
\cite{Mussardo:1998kq}-\cite{HSG2}).

So far, we have briefly recalled the program for bulk integrable
QFTs, that is QFTs without boundaries or defects. Given that most
realistic physical systems possess boundaries and/or defects, it
is both interesting and natural to ask the question whether or not
the integrability of certain 1+1 dimensional QFTs may be preserved
in the presence of boundaries or defects. The study of the
conditions under which that question can be answered in the
affirmative has attracted a lot of attention in the last two
decades, with many works devoted to the study of integrable QFTs
with boundaries \cite{Cherednik:1985vs}-\cite{Bowcock:1995vp} and
defects \cite{Delfino:1994nr,impurityus,annecy}. A natural further
step is trying to extend the bulk form factor program just
recalled to boundary and defect theories. Here we want to
concentrate on the boundary case. Given an integrable QFT with a
boundary there are two possible approaches to the problem of
computing correlation functions:
\paragraph{Boundary at $t=0$:} we may place the boundary at $t=0$ extending along the
    space direction ($t$ represents  the time coordinate). In this case it is natural to characterize the presence of the boundary by means of a boundary state $|B\rangle$.
    The Hilbert space is exactly the same as for the bulk
    theory, and therefore form factors can be completely
    characterized by those in the bulk. The only additional
    information needed is an explicit description of the boundary
    state, which has been provided by S.~Ghoshal and A.B.~Zamolodchikov
    \cite{Ghoshal:1993tm},
    \begin{equation}\label{bstate}
    |B\rangle:=\exp\left(\frac{1}{4\pi}\int_{-\infty}^{\infty}R_{\mu_i}(\frac{i\pi}{2}-\theta)Z_{\mu_i}(-\theta)Z_{\mu_i}(\theta)d\theta\right)
    |0\rangle.
    \end{equation}
   Here $R_{\mu_i}(\theta)$ is the reflection amplitude off the
    boundary, which we take to be diagonal in this case (particles just reflect off the boundary with
    a certain probability, without changing their species).
Expanding the exponential above, boundary form factors can be
expressed in terms of bulk form factors, although the former will
then be an infinite sum of the latter. This sum is in general very
involved, but can be performed analytically for some correlation
functions of free theories \cite{Konik:1995ws}. This picture can
be generalized to the defect case, by replacing the boundary state
above by a defect state. A realization of this state and explicit
form factor computations have been carried out for free theories
\cite{Delfino:1994nr,impurityus} with different types of defects.
More recently it has been proven in  \cite{Bajnok:2006} that a
defect theory may always be regarded alternatively as a boundary
theory by means of the so-called ``folding trick". Exploiting this
correspondence, the authors also constructed the defect state for
integrable models with purely transmitting defects.
\paragraph{Boundary at $x=0$:} we may  place the boundary
at $x=0$ extending in the time direction ($x$ represents the
position coordinate). In this case the Hilbert space will change
with respect to the bulk case, as only half of the physical states
remain linearly independent in the presence of a boundary. This is
so since in addition to (\ref{Zalg1})-(\ref{Zalg3}) the further
constraint
\begin{equation}\label{ind}
   Z_{\mu_{i}}(\theta_i)= R_{\mu_i}(\theta_i)
   Z_{\mu_{i}}(-\theta_i),
\end{equation}
holds. Recently, a generalization of the bulk form factor program
for theories with boundaries and for boundary operators (that is,
operators defined on the boundary) has been proposed \cite{BPT},
and its validity tested against previously known results for
various QFTs. It has been established that the relations
(\ref{Zalg1})-(\ref{Zalg3}) and (\ref{ind}) together with the
analytical properties of $S$- and $R$-matrices lead to a set of
consistency equations for the boundary form factors, which can be
solved for particular models along similar lines as for bulk
theories. Some of these solutions were found in \cite{BPT},
although in most cases only for low particle numbers. The main aim
of this paper is to extend some of the results in \cite{BPT} by
carrying out a detailed study of the boundary form factor
solutions of the sinh-Gordon model with a particular type of
Dirichlet boundary conditions (corresponding to vanishing value of
the sinh-Gordon field at the boundary) at the self-dual point.

The paper is organized as follows: In section 1 we review the
boundary form factor program as proposed in \cite{BPT}. In section
2 we recall the $S$- and $R$-matrices of the sinh-Gordon model
with Dirichlet boundary conditions and vanishing boundary field.
We also present the form factor equations found in \cite{BPT} for
this particular model. In section 3 we re-write these equations in
terms of elementary symmetric polynomials and analyze their
special properties at the self-dual point.  In section 4 we
provide new explicit solutions to the form factor equations up to
16-particles and formulate a conjecture for the form of all
$n$-particle form factors of the boundary operators
$\partial_{x}\phi$ and $(\partial_{x} \phi)^2$. In section 5 we
find solutions to the form factor consistency equations for other
boundary fields. We summarize our conclusions in section 6.
Finally, we list various explicit form factor formulae in the
appendix.
\section{The boundary form factor program}
In this section we will review the main aspects of the boundary
form factor program proposed in \cite{BPT}, for diagonal
scattering theories possessing a single particle spectrum and no
particle bound states (such as the sinh-Gordon model). For such a
model, the particle indices $\mu_i$ in (\ref{ff}) can be dropped
and the boundary form factor axioms written as
\begin{eqnarray}
  F_{n}^{\mathcal{O}}(\theta_1, \ldots,\theta_i, \theta_{i+1}, \ldots \theta_n) &=&
  S(\theta_{i\,i+1})
  F_{n}^{\mathcal{O}}(\theta_1, \ldots,\theta_{i+1}, \theta_i,  \ldots \theta_n), \label{cpt}\\
 F_{n}^{\mathcal{O}}(\theta_1, \ldots, \theta_{n-1},
\theta_n) &=&
  R_{n}(\theta_n)
  F_{n}^{\mathcal{O}}(\theta_1, \ldots, \theta_{n-1},
  -\theta_n),\\
 F_{n}^{\mathcal{O}}(\theta_1,\theta_2, \ldots,
\theta_n) &=&
  R_{n}(i\pi-\theta_1)
  F_{n}^{\mathcal{O}}(2 \pi i-\theta_1,\theta_2, \ldots,\theta_n),
  \label{crossing}
\end{eqnarray}
The first two axioms follow from the braiding relations
(\ref{Zalg1}) and (\ref{ind}), whereas the last axiom expresses
crossing symmetry in the presence of a boundary. In addition we
have the kinematical residue equations which generalize the
corresponding ones in the bulk
\begin{eqnarray}
 -i \text{Res}_{\substack{\bar{\theta}_{0}={\theta}_{0}}}
 F_{n+2}^{\mathcal{O}}(\theta_0+i\pi,\bar{\theta}_{0}, \theta_1 \ldots, \theta_n)
  =\left(1-\prod_{i=1}^{n} S(\theta_{0i}) S(\hat{\theta}_{0i})\right)
  F_{n}^{\mathcal{O}}(\theta_1, \ldots,\theta_n),\label{kre}
\end{eqnarray}
with $\hat{\theta}_{ij}:=\theta_i+\theta_j$. It was argued in
\cite{BPT} that a second kinematical relation which has no
counterpart in the bulk form factor program emerges whenever the
reflection amplitude $R(\theta)$ has a pole at $\theta=i\pi/2$. We
will not report this equation here, since it is trivial for the
particular model we want to study later.

As in the bulk case, further recursive relations exist in the
presence of particle bound states but we will not review them
here. Mimicking the solution procedure developed for the bulk case
and diagonal scattering matrices, an ansatz for the solutions of
the above equations can be made, which solves axioms
(\ref{cpt})-(\ref{crossing}) by construction and at the same time
makes explicit the form factor pole structure
\begin{equation}\label{ansatz}
    F_{n}^{\mathcal{O}}(\theta_1, \ldots, \theta_n)=H^{\mathcal{O}}_n Q^{\mathcal{O}}_n(y_1, \ldots,
    y_n)\prod_{i=1}^{n} {r(\theta_i) }\prod_{1\leq i<j \leq n} \frac{f(\theta_{ij})f(\hat{\theta}_{ij})}{y_i +
    y_j}.
\end{equation}
Here $y_i=2\cosh \theta_i$, $H^{\mathcal{O}}_n$ is a constant and
$Q^{\mathcal{O}}_n(y_1, \ldots, y_n)$ is an entire function of the
variables $y_1,\ldots, y_n$. The function $f(\theta)$ is the same
minimal (2-particle) form factor found in the bulk case and
$r(\theta)$ is the minimal 1-particle form factor. Integral
representations for these minimal form factors can be readily
obtained from the representations of the $S$-matrix and reflection
amplitude $R(\theta)$, respectively.
\section{Boundary form factors of the sinh-Gordon model with
Dirichlet boundary conditions} The particular example we want to
study in this letter is the sinh-Gordon theory with a particular
type of Dirichlet boundary conditions. The sinh-Gordon model is
one of the simplest integrable models possessing a non-trivial
scattering matrix since its spectrum consists of a single particle
and no particle bound states occur. The bulk scattering matrix
\cite{SSG2,SSG3,SSG} is a pure CDD factor given by
\begin{equation}\label{smatrix}
    S(\theta)=-(-B)_{\theta}(B-2)_{\theta},
\end{equation}
in terms of the blocks
\begin{equation}\label{block}
(x)_{\theta}=\frac{\sinh\frac{1}{2}\left(\theta + \frac{i \pi
x}{2}\right)} {\sinh\frac{1}{2}\left(\theta - \frac{i \pi
x}{2}\right)}.
\end{equation}
The parameter $B \in [0,2]$ is the effective coupling constant
which is related to the coupling constant $\beta$ in the
sinh-Gordon Lagrangian \cite{toda2,toda1}
\begin{equation}\label{lagran}
    \mathcal{L}=\frac{1}{2}(\partial_{\mu}\phi)^2-\frac{m^2}{\beta^2}\cosh(\beta\phi),
\end{equation}
where $m$ is a mass scale and $\phi$ is the sinh-Gordon field, as
\begin{equation}\label{BB}
    B(\beta)=\frac{2\beta^2}{8\pi + \beta^2},
\end{equation}
under CFT normalization \cite{za}. The $S$-matrix is obviously
invariant under the transformation $B\rightarrow 2-B$, a symmetry
which is also referred to as week-strong coupling duality, as it
corresponds to $B(\beta)\rightarrow B(\beta^{-1})$ in (\ref{BB}).
The point $B=1$ is known as the self-dual point. When an
integrable boundary is introduced and Dirichlet boundary
conditions imposed the corresponding reflection amplitude
$R(\theta)$ was found to be \cite{CT}
\begin{equation}
    R(\theta)=\left(1\right)_{\theta}\left(1+\frac{B}{2}\right)_{\theta}\left(2-\frac{B}{2}\right)_{\theta}
    \frac{\left({E-1}\right)_{\theta}}{(E+1)_{\theta}}.\label{r}
\end{equation}
This expression was obtained by analytical continuation of the
sine-Gordon breather reflection amplitudes calculated in
\cite{Ghoshal:1993iq}. In (\ref{r}) $E$ is a continuous parameter
which is proportional to the boundary value of the sinh-Gordon
field as
\begin{equation}\label{E}
    E= \frac{4 i B \phi(0)}{\beta}.
\end{equation}
At the special value $E=\phi(0)=0$ which we will consider in this
manuscript, the reflection amplitude (\ref{r}) has no pole at
$\theta=i\pi/2$ and therefore the ansatz (\ref{ansatz}) captures
the full pole structure of the boundary form factors. In addition,
since Dirichlet boundary conditions imply a fixed value of
$\phi(0)$, the boundary operator content of the theory consists
solely of products of space-derivatives of the boundary field like
$\partial_{x}\phi$, $(\partial_{x} \phi)^2$, that is, derivatives
in the direction perpendicular to the boundary. The recursive
equation obtained by substituting (\ref{ansatz})  in the kinematic
residue equation (\ref{kre}) was derived in \cite{BPT} and takes
the form
\begin{equation}\label{Q}
    Q^{\mathcal{O}}_{n+2}(-y,y,y_1, \ldots, y_n)=-P_n (y,y_1,\ldots,y_n) Q^{\mathcal{O}}_n(y_1, \ldots, y_n).
\end{equation}
The polynomial $P_n$ was also given in \cite{BPT}. Here we present
it in a slightly different form, in terms of elementary symmetric
polynomials in the variables $y_1, \ldots, y_n$,
\begin{equation}\label{Pn}
    P_n(y,y_1,\ldots,y_n)=\frac{i (-\omega_{+}\omega_{-})^{n}}{\omega_{+}-\omega_{-}}\sum_{k,p=0}^{n}
    (-i \omega_{-})^{-k} (-i \omega_{+})^{-p} \, \sigma_{k}^{n} \sigma_{p}^{n} \sin(\pi
    (k-p)/{2}),
\end{equation}
which are defined through the generating function\footnote{In what
follows we will employ the simpler notation
$\sigma_{k}^{n}=\sigma_{k}$, since the number of variables $n$ on
which the polynomials depend is usually obvious from the context
in which they occur. }
\begin{equation}\label{esp1}
\prod_{k=1}^{n}(z + y_k)= \sum_{k=0}^{n} z^{n-k} \sigma_{k}^{n},
\end{equation}
The variables $\omega_{\pm}$ are given by
\begin{equation}\label{ypm}
\omega_{\pm}=2 \cosh(\theta\pm i \pi B/2).
\end{equation}
\subsection{Recursive relations at the self-dual point}\indent \\

\noindent In this section we initiate the presentation and
discussion of the new results obtained in this manuscript. Whereas
in \cite{BPT} computations up to $n=5$ particle form factors and
for generic values of $B$ where carried out, here we have
concentrated our analysis on the special case $B=1$, that is the
self-dual point. In this section we will show how at this point a
new factorized structure is found, both of the form factor
equations and their solutions, which will allow us to find form
factor solutions up to much higher particle numbers ($n=16$) and
to put forward a plausible hypothesis for the structure of the
solutions of the form factor equations for arbitrary particle
numbers. Let us start by noticing that for $B=1$ the variables
$\omega_{\pm}$ become proportional to each other
\begin{equation}
\omega_{\pm}=\pm 2 i \sinh \theta,
\end{equation}
and this allows us after some simple algebra to bring $P_n$ into
the factorized form
\begin{equation}\label{Pn2}
P_n(y,y_1,\ldots,y_n)=  P^{e}_n(y,y_1,\ldots,y_n)
P^{o}_n(y,y_1,\ldots,y_n),
\end{equation}
with
\begin{eqnarray}
P^{e}_n(y,y_1,\ldots,y_n) &=& (-i
\omega_{+})^{2\left[\frac{n}{2}\right]}  \sum_{p=0}^{n}
\sigma_{p}^{n} (-i \omega_{+})^{-p} \, \cos(\pi p/2),
\nonumber \\
P^{o}_n(y,y_1,\ldots,y_n) &=& (-1)^{n+1}(-i \omega_{+})^{2
\left[\frac{n+1}{2}\right] -1} \sum_{p=0}^{n} \sigma_{p}^{n} (-i
\omega_{+})^{-p} \, \sin(\pi p/2),\label{Peo}
\end{eqnarray}
where $[x]$ denotes the integer part of $x$. Clearly, the
polynomials $P^{e}_n$ and $P^{o}_n$ involve only elementary
symmetric polynomials of even and odd degree, respectively. In
addition, the coefficients of the symmetric polynomials in
$P^{e}_n$ and $P^{o}_n$ are always even powers of $-i \omega_{+}$,
that is, $P^{e}_n$ and $P^{o}_n$ are polynomials in $(i
\omega_{+})^{2}=y^{2}-4$, of which the first few examples are
\begin{equation}
    \begin{array}{rclrcl}
      P^{e}_1 &=& 1, & P^{o}_1 &= &\sigma_{1}, \\
      P^{e}_2 &=& (i \omega_{+})^{2}-\sigma_{2},&P^{o}_2&=& -\sigma_{1},\\
      P^{e}_3&=& (i \omega_{+})^{2}-\sigma_{2},& P^{o}_3&=&(i \omega_{+})^{2}\sigma_{1}-\sigma_{3},\\
      P^{e}_4 &=& (i \omega_{+})^{4}-(i \omega_{+})^{2}\sigma_{2}+
\sigma_{4},& P^{o}_4&=&-(i
\omega_{+})^{2}\sigma_{1}+\sigma_{3}.\\
    \end{array}
\end{equation}
It was noticed in \cite{BPT} that for $E=0$ the symmetry of the
Lagrangian under $\phi \rightarrow -\phi$, where $\phi$ is the
sinh-Gordon fundamental field, is preserved in the boundary
theory. This means that, depending on the operator $\mathcal{O}$
considered, only the even or odd particle form factors will be
non-vanishing, exactly as in the bulk theory
\cite{FMS,KK,Delfino:2006te}.

In \cite{BPT} the first non-vanishing form factors of two
operators identified as $\partial_{x} \phi$ and $(\partial_x
\phi)^2$ were computed. Setting $B=1$, the solutions found become
\begin{eqnarray}
Q_1^{\partial_{x} \phi}&=& 1,\\
Q_2^{(\partial_{x} \phi)^2}&=& -\sigma_{1},\\
Q_3^{\partial_{x} \phi}&=&-\sigma_{1},\\
Q_4^{(\partial_{x} \phi)^2}&=& \sigma_{1}^{2}(4+\sigma_{2}),\\
Q_5^{\partial_{x} \phi}&=& \sigma_{1} (4\sigma_{1} + \sigma_{3})(4
+ \sigma_{2}).
\end{eqnarray}
Here we have presented the solutions for both operators in order
of increasing particle number to make certain features more
apparent: first of all, all $Q_n$ factorize into a factor
containing only elementary symmetric polynomials of even degree
and another factor containing only those of odd degree. Secondly,
some of these factors repeat in various form factors (like
$4+\sigma_2$ which appears both in $Q_4^{(\partial_{x} \phi)^2}$
and $Q_5^{\partial_{x} \phi}$), so that some sort of pattern seems
to emerge. In this letter strong evidence will be provided that
these two features are actually general: all functions
$Q_n^{\partial \phi}, Q_n^{(\partial_{x} \phi)^2}$ factorize and a
general pattern for the $n$-particle solutions exists, such that
once the form factors of $\partial_{x} \phi$ are known also those
of $(\partial_{x} \phi)^2$ are automatically determined and
viceversa.

Recalling that the $P$-function (\ref{Pn}) factorizes in a similar
way, our results suggest that the recursive equations (\ref{Q})
can be split up into two simpler recursive equations involving
elementary symmetric polynomials of even and odd degree,
respectively
\begin{equation}\label{QQ}
    Q_n^{\mathcal{O}}(y_1,\ldots,y_n)=Q_n^{\mathcal{O}|e}(y_1,\ldots,y_n)
    Q_n^{\mathcal{O}|o}(y_1,\ldots,y_n),
\end{equation}
with
\begin{eqnarray}
    Q_{n+2}^{\mathcal{O}|e}(-y,y,y_1,\ldots,y_n)&=&
    P^{e}_n(y,y_1,\ldots,y_n)Q_n^{\mathcal{O}|e}(y_1,\ldots,y_n),\nonumber \\
    Q_{n+2}^{\mathcal{O}|o}(-y,y,y_1,\ldots,y_n)&=&-
    P^{o}_n(y,y_1,\ldots,y_n)Q_n^{\mathcal{O}|o}(y_1,\ldots,y_n).\label{Qeo}
\end{eqnarray}
In this new notation, the solutions above become:
\begin{equation}
    \begin{array}{rclrcl}
      Q_1^{\partial_{x} \phi|e} &=& 1, & Q_1^{\partial_{x} \phi|o}&=&1, \\
      Q_2^{(\partial_{x} \phi)^2|e}&=& -1,& Q_2^{(\partial_{x} \phi)^2|o}&=& \sigma_{1},\\
      Q_3^{\partial_{x} \phi|e}&=&1,& Q_3^{\partial_{x} \phi|o}&=& -\sigma_{1},\\
     Q_4^{(\partial_{x} \phi)^2|e}&=& 4+\sigma_{2},&Q_4^{(\partial_{x} \phi)^2|o}&=& \sigma_{1}^{2},\\
  Q_5^{\partial_{x} \phi|e}&=& -(4 + \sigma_{2}),& Q_5^{\partial_{x} \phi|o}&=&-\sigma_{1} (4\sigma_{1} +
\sigma_{3}).\label{e1e4}
    \end{array}
\end{equation}
This factorization simplifies greatly the process of finding
solutions of (\ref{Q}) for higher particle numbers. From
(\ref{Qeo}) and (\ref{Peo}) two important characteristics of the
solutions of (\ref{Qeo}) can be deduced:
\begin{itemize}
    \item the functions $Q_{2n}^{\partial_{x} \phi|e}$
and $Q_{2n+1}^{(\partial_{x} \phi)^2|e}$ always coincide, but for
the number of variables entering the symmetric polynomials and a
sign. The reason is that $P_{2n}^{e}$ and $P_{2n+1}^{e}$ are in
fact the same in terms of elementary symmetric polynomials and
that the input values $Q_{1}^{(\partial_{x}
\phi)^2|e}=-Q_{2}^{\partial \phi|e}=1$ are the same but for a
sign.
    \item Similarly, $Q_{2n+1}^{(\partial_{x} \phi)^2|o}$
and $Q_{2n+2}^{\partial_{x} \phi|o}$  also coincide but for an
extra $-\sigma_1$ factor in $Q_{2n+2}^{\partial_{x} \phi|o}$. This
is again due to the fact that the corresponding $P$-functions,
$P^{o}_{2n+1}$ and $P^{o}_{2n+2}$ are identical up to a sign in
terms of elementary symmetric polynomials and that the initial
values $Q_{2}^{\partial_{x} \phi|o}= \sigma_1$ and
$Q_{1}^{(\partial_{x} \phi)^2|o}=1$ differ only by a $\sigma_1$
factor which satisfies the first equation in (\ref{esp}).
\end{itemize}
These features can be found on the few examples listed above. Most
importantly, they allow us to conclude that, at the self-dual
point, the knowledge of all non-vanishing form factors of the
operator $(\partial_x \phi)^2$ automatically implies the knowledge
of all non-vanishing form factors of the operator  $\partial_{x}
\phi$ through the relations
\begin{eqnarray}\label{rr1}
    Q_{2n}^{(\partial_{x} \phi)^2|e}&=&-Q_{2n+1}^{\partial_{x} \phi|e},\quad \text{for}\quad
    n \in \mathbb{Z}^{+},\\
    Q_{2n}^{(\partial_{x} \phi)^2|o}&=&-\sigma_1 Q_{2n-1}^{\partial_{x} \phi|o},\quad \text{for}\quad
    n \in \mathbb{Z}^{+}.\label{rr2}
\end{eqnarray}
Note that these equalities are only true when the functions are
expressed in terms of elementary symmetric polynomials. The latter
will obviously depend on different particle numbers on the r.h.s.
and l.h.s. of the equations (and therefore, on different numbers
of variables).

\section{Solutions for higher particle numbers}
We shall now attempt to find new solutions to (\ref{Qeo}) in terms
of elementary symmetric polynomials in the standard way employed
previously for several bulk theories (see e.g.
\cite{FMS,KK,Z,HSG1,HSG2}). This solution procedure can be rather
tedious as the number of terms on the r.h.s. of (\ref{Qeo})
increases rapidly with the particle number. From these terms one
must be able to ``reconstruct" the l.h.s. of the equation when the
two first variables are not $y$ and $-y$ but generic. This
reconstruction can be performed by employing the following
relations between elementary symmetric polynomials depending on
$n+2$ and $n$ variables
\begin{eqnarray}\label{esp}
 \sigma_{k}(-y,y,y_1,\ldots, y_n)&=&-y^{2}\sigma_{k-2}(y_1,\ldots, y_n)+ \sigma_{k}(y_1,\ldots, y_n),
\end{eqnarray}
with $\sigma_{-1}=0$ and $\sigma_{0}=1$. The relations (\ref{esp})
allow us to make a guess about the possible origin of each of the
terms on the r.h.s. of (\ref{Qeo}). For example, if a term
$\sigma_2^{n}$ appears it can only originate from a term
$\sigma_2^{n+2}$ in the l.h.s. and if a term $-y^2 \sigma_n^{n}$
appears, then it can originate either from a $\sigma_{n+2}^{n+2}$
or a $\sigma_2^{n+2} \sigma_n^{n+2}$ term on the r.h.s. By
consistently accounting for each term on the r.h.s. of (\ref{Qeo})
a unique polynomial solution to the equations can be found.
Solving the recursive equations systematically in this way we have
found explicit solutions up to 16-particle form factors. All the
solutions found exhibit the following general structure:
\begin{eqnarray}\label{gf}
    Q_{2n}^{(\partial_{x} \phi)^2|e}&=&(-1)^{n+1}Q_{2n-2}^{(\partial_{x} \phi)^2|e}\mu_{2n-2} +
    \sigma_{2n}R_{2n}^{e},\\
    Q_{2n}^{(\partial_{x} \phi)^2|o} & =& (-1)^n Q_{2n-2}^{(\partial_x
\phi)^2|o}\mu_{2n-3} +  \sigma_1^{2}\sigma_{2n-1}
    R_{2n}^{o}, \label{gf2}
\end{eqnarray}
Let us now explain the various terms appearing in
(\ref{gf})-(\ref{gf2}):
\begin{itemize}
    \item The first term on the r.h.s. of the equations shows a recursive
relation between $2n$- and $(2n-2)$-particle solutions of
(\ref{Qeo}). However, the polynomials $Q_{2n-2}^{(\partial_x
\phi)^2|e}$ and $Q_{2n-2}^{(\partial_{x} \phi)^2|o}$ appearing on
the r.h.s. of the above equations are not any more the solutions
of (\ref{Qeo}). They are identical to them in terms of elementary
symmetric polynomials, but these are now polynomials on all $2n$
variables involved on the l.h.s. of the equation.
    \item The functions $\mu_i$ are defined as follows:
\begin{eqnarray}\label{pp}
\mu_0&=&1, \qquad \mu_1= \sigma_1, \nonumber \\ \mu_{i}&=&4
\mu_{i-2}+ \sigma_{i}, \quad \text{for}\quad i>1,
\end{eqnarray}
in terms of elementary symmetric polynomials. Equivalently
\begin{equation}
\mu_{i} = \sum_{k=0}^{[i/2]} 4^{k} \sigma_{i-2k},
\end{equation}
 where, as before $[x]$ stands for the integer part of $x$. The
 emergence of such particular combinations of elementary symmetric
 polynomials is rather natural given their reduction properties,
 which can be deduced from (\ref{esp}):
\begin{eqnarray}\label{esp2}
 \mu_{k}(-y,y,y_1,\ldots,y_n)&=&(4-y^{2})\mu_{k-2}(y_1,\ldots,y_n)+ \sigma_{k}(y_1,\ldots,y_n),
\end{eqnarray}
where $\mu_{-1}=0$ and $\mu_0=1$. As we can see the variable $y$
appears always in the combination $4-y^2$, which is precisely the
same variable entering the polynomials $P^{e}$ and $P^{o}$.
 \item Further, the conjectured expressions
(\ref{gf})-(\ref{gf2}) involve the additional terms $\sigma_{2n}
R_{2n}^{e}$ and
    $\sigma_1^{2}\sigma_{2n-1} R_{2n}^{o}$, respectively. One can easily convince
    oneself from the previous definitions, that these are in fact
    the only terms containing the elementary symmetric polynomials $\sigma_{2n}$
    and $\sigma_{2n-1}$, respectively.
    Therefore $\sigma_{2n}R_{2n}^{e}$ and $\sigma_1^2 \sigma_{2n-1}R_{2n}^{o}$
    are the terms containing the elementary symmetric polynomial of
     maximum degree.
    This definition makes it straightforward to reconstruct the
    functions $R_{2n}^{e}$ and $R_{2n}^{o}$ for particular cases
    up to high particle numbers.
From (\ref{gf}), (\ref{gf2}), (\ref{Qeo}), (\ref{esp}) and
(\ref{esp2}) it
    follows that they must satisfy the following reduction properties:
   \begin{eqnarray}\label{gf3}
    && -y^2  R_{2n}^{e}(-y,y,y_1,\ldots,y_{2n-2})\sigma^{2n-2}_{2n-2}=
    P_{2n-2}^{e}Q_{2n-2}^{(\partial_{x} \phi)^2|e}(y_1,\ldots,y_{2n-2})\nonumber\\
    && +(-1)^{n}Q_{2n-2}^{(\partial_{x} \phi)^2|e}(-y,y,y_1,\ldots,y_{2n-2})
    ((4-y^2)\mu_{2n-4}^{2n-2}+\sigma_{2n-2}^{2n-2}),\\
  && y^2  (\sigma_1^{2n-2})^{2} R_{2n}^{o}(-y,y,y_1,\ldots,y_{2n-2})\sigma^{2n-2}_{2n-3}=
    P_{2n-2}^{o}Q_{2n-2}^{(\partial_{x} \phi)^2|o}(y_1,\ldots,y_{2n-2})\nonumber\\
    && +(-1)^{n}Q_{2n-2}^{(\partial_{x} \phi)^2|o}(-y,y,y_1,\ldots,y_{2n-2})
    ((4-y^2)\mu_{2n-5}^{2n-2}+\sigma_{2n-3}^{2n-2}),\label{gf4}
\end{eqnarray}
where, in order to avoid confusion, we have made the variable
dependence explicit and
\begin{equation}
    \sigma_{k}^{2n-2}=  \sigma_{k}(y_1,\ldots,y_{2n-2})
    \quad \text{and}\quad \mu_{k}^{2n-2}=
    \mu_{k}(y_1,\ldots,y_{2n-2}),
\end{equation}
for any values of $k$. Notice that, once the conjectured formulae
    (\ref{gf})-(\ref{gf2}) are assumed to hold, $R_{2n}^{e}$ and
    $R_{2n}^{o}$ are the only unknowns, which means an
    enormous simplification of the original problem. Explicit formulae for
    $R_{2n}^{e}$ and $R_{2n}^{o}$ up to $n=8$ can be found in the appendix.
    \end{itemize}
In addition to the structure just outlined, a simple recipe allows
to relate the solutions $Q_{2n}^{(\partial_{x} \phi)^2|e}$ and
$Q_{2n+2}^{(\partial_{x} \phi)^2|o}$. The latter can always be
obtained from the former by performing the replacements $\sigma_k
\rightarrow \sigma_{k+1}$ and $\mu_k \rightarrow \mu_{k+1}$ and
multiplying by a global factor $-\sigma^{2}_1$. Once these
transformations have been done, we must still introduce as many
factors $\sigma_1$ as to achieve that each term in
$Q_{2n+2}^{(\partial_{x} \phi)^2|o}$ is exactly a product of $n+1$
symmetric polynomials. For example, if a term $\sigma_8$ would
appear in $Q_{8}^{(\partial_{x} \phi)^2|e}$, then it would imply a
term $-\sigma_1^2\sigma_1\sigma_9$ in $Q_{10}^{(\partial_x
\phi)^2|e}$. Thus, provided we know $ Q_{2n}^{(\partial_{x}
\phi)^2|e}$ we can systematically determine $Q_{2n}^{(\partial_x
\phi)^2|o}$. From these solutions, as concluded above
(\ref{rr1})-(\ref{rr2}), also all functions
$Q_{2n+1}^{\partial_{x} \phi|e}$ and $Q_{2n+1}^{\partial_{x}
\phi|e}$ can be obtained.
\section{Solutions for other boundary fields}
In this section we will address the problem of finding solutions
of the form factor equations for fields other than  $\partial_x
\phi$ and $(\partial_x \phi)^{2}$. It is worth recalling that the
fields $\partial_x \phi$ and $(\partial_x \phi)^{2}$ were
identified in \cite{BPT} by analyzing the asymptotic behaviour of
their form factors. Although this identification seems plausible,
it would be very interesting to strengthen it with some numerical
work.

A particular way of looking for new solutions of the form factor
equations is to find entire functions $I_{n}^{s}(y_1,\ldots,y_n)$
such that if $F_{n}^{\mathcal{O}}(y_1,\ldots,y_n)$ is a solution,
then the product
\begin{equation}\label{if}
    F_{n}^{\mathcal{O}'}(y_1,\ldots,y_n)=I^{s}_{n}(y_1,\ldots,y_n)
    F_{n}^{\mathcal{O}}(y_1,\ldots,y_n),
\end{equation}
is also one. In the bulk case, such type of form factors (with
variables $y_i=e^{\theta_i}+e^{-\theta_i}$ replaced by
$x_i=e^{\theta_i}$) have been related to ``descendent" fields for
the first time in the work of J.~L.~Cardy and G.~Mussardo
\cite{Cardybast} for the Ising model. The name descendent must be
understood in the sense that if $\mathcal{O}$ is related to a
spinless primary field of the underlying CFT in the ultraviolet
limit, then $\mathcal{O}'$ would be related to a descendent field
of $\mathcal{O}$ in the same limit. Which particular class of
descendent fields $\mathcal{O}'$ relates to is determined by the
scaling properties of the function $I_n^{s}$ under a rapidity
shift, that is, the spin of $\mathcal{O}'$ which we call $s$.
Solutions to the bulk form factor equations of the type (\ref{if})
have been also found for the sinh-Gordon \cite{FMS} and  Yang-Lee
\cite{Christe:1990zt} models.

In our case, the link between form factors of the type (\ref{if})
and descendent fields in the conformal limit is a priory less
clear, as the operator content of the CFT at the boundary is not
classified in terms of spin representations. However, we may still
attempt to find solutions of the form (\ref{if}). They will be
related to new fields of the QFT, that is higher derivatives of
$\phi$ which is very natural to relate to descendent fields at
conformal level. In this case the superscript $s$ in $I_{n}^{s}$
is just the leading order behaviour of $I_n^s$ when all rapidities
tend to infinity. The analysis is in fact completely analog to the
one presented in \cite{FMS} for the bulk sinh-Gordon model. We
just need to replace the variables $x_i$ by the variables $y_i$.
First of all, it is trivial to see that the form factor (\ref{if})
will automatically satisfy all consistency equations
(\ref{cpt})-(\ref{crossing}) provided $I_{n}^{s}(y_1,\ldots,y_n)$
can be entirely expressed in terms of elementary symmetric
polynomials in the variables $\{y_1,\ldots,y_n\}$. Secondly, if
(\ref{if}) is to satisfy the kinematic residue equation
(\ref{kre}) then $I_{n}^{s}$ must satisfy the following reduction
property:
\begin{equation}\label{Is}
    I_{n+2}^{s}(-y,y,y_1,\ldots,y_n)=I^{s}_{n}(y_1,\ldots,y_n).
\end{equation}
This is exactly the same equation found in \cite{Cardybast,FMS}. A
basis of the space of solutions to (\ref{Is}) has been found to be
\begin{equation}\label{Idet}
    I_{n}^{2s-1}=(-1)^{s+1}\det\mathcal{I},
\end{equation}
with $\mathcal{I}$ being a matrix of entries
\begin{equation}\label{II}
   \mathcal{I}_{1j}=\sigma_{2j-1} \qquad\text{and}\qquad
   \mathcal{I}_{ij}=\sigma_{2j-2i+2},
\end{equation}
for $j=1,\ldots,s$ and $i=2,\ldots,s$. As in the bulk case
$\det\mathcal{I}$ is of order $2s-1$ in terms of elementary
symmetric polynomials.

To finish this section it is worth mentioning that it would be
very interesting to establish rigorously a correspondence between
the number of solutions to the form factor equations for fixed $s$
and the number of descendent fields of the underlying CFT at level
$s$. For this one should proceed along the lines of the analysis
carried out in \cite{Cardybast} for the bulk Ising model. It is
however to be expected that such analysis becomes more involved
for the present model.
\section{Conclusions and outlook}
In this paper we have studied the form factors of two boundary
operators of the sinh-Gordon model with a particular type of
Dirichlet boundary conditions, corresponding to vanishing boundary
value of the sinh-Gordon field, at the self-dual point. We have
found that both the form factor recursive equations (\ref{Q}) and
their solutions exhibit distinct properties at $B=1$: they both
factorize into one factor containing only elementary symmetric
polynomials of even degree and another term containing only those
of odd degree. This factorization has allowed us to obtain
explicit solutions of the equations up to remarkably high particle
numbers and to provide a conjecture for the structure of the form
factor solutions for arbitrary particle numbers. Unfortunately we
have not yet been able to provide a mathematically rigorous proof
of that structure, although the many examples computed strongly
support it.

The finding of closed formulae for all $n$-particle form factors
of local fields is a highly complicated task which has only been
successfully carried out for few bulk scattering theories and
operators \cite{Z,Fring:1992pj,FMS,KK,Delfino:1994ea,HSG1,HSG2}.
Even when closed formulae have been found, rigorous proofs of
their general validity have only been given in some cases
\cite{FMS,HSG1,HSG2}. A crucial element of these proofs has been
the fact that the form factor solutions were given in terms of
determinants whose entries were elementary symmetric polynomials.
It is this determinant structure and the possibility of employing
general properties of determinants which made these proofs
possible. The formulae conjectured in the present paper
(\ref{gf})-(\ref{gf2}) are however not of determinant form, so
that general proofs should go along very different lines. In
addition, it is worth noticing an important difference with
respect to the bulk form factor program with regard to the
structure of the form factor solutions: boundary fields do not
possess a definite spin (since translation invariance is broken)
and therefore the transformation under a rapidity shift of the
allowed terms in the form factor formulae is not constrained. This
means that not all terms in the functions $Q_{n}$ need to have the
same degree. As a result, formulae tend to be more involved and
the form factor's general structure is harder to identify. It
would be very interesting to investigate whether or not some
structures of the boundary form factors may be universal or at
least common to many models. By this we mean structures such as
the determinant formulae mentioned above for bulk theories. For
the case at hand, it would be of course desirable to find a
general pattern for the functions $R_{n}^{e}, R_{n}^{o}$, which
may also facilitate proofs.

In addition, new solutions to the form factor consistency
equations related to local operators other than $\partial_x \phi$
and $(\partial_x \phi)^2$ have been found. We have seen that the
form factors of these fields are related to the ones of
$\partial_x \phi$ and $(\partial_x \phi)^2$ by a multiplicative
factor (\ref{if}). The same kind of structure has been previously
identified for the form factors of the bulk Ising
\cite{Cardybast}, sinh-Gordon \cite{FMS} and Yang-Lee models
\cite{Christe:1990zt} and has been related to descendant fields at
the conformal level. It is natural to assume that this
interpretation also holds here, however two interesting open
problems remain: first, a numerical study of the short-distance
behaviour of correlation functions of these fields should be
carried out in order to make the identification with descendents
at conformal level more clear. Second, a rigorous study of the
correspondence between form factor solutions with a particular
asymptotic behaviour and descendent fields for fixed level should
be carried out along the lines of \cite{Cardybast}.

We would like to conclude by saying that the program proposed in
\cite{BPT} is very recent and has so far only been pursued for the
few models treated in the original paper and for the particular
case considered here in more detail. A lot of work remains to be
done to develop the boundary form factor program to a degree
comparable to that achieved for bulk theories. Obviously computing
boundary form factors of other integrable models would be
interesting, particularly so for non-diagonal theories such as the
sine-Gordon model. For the latter theory boundary form factors
have in fact been computed in \cite{boyuhu} by a different
approach but it would still be interesting to employ the method of
\cite{BPT} for the same computation as a consistency check and/or
as a possible way to obtain alternative (maybe more explicit)
representations of the form factors. It would also be very
interesting to exploit these form factors solutions in similar
ways as has been done for bulk theories: computing correlation
functions, investigating the operator content and computing
characteristic quantities of the underlying boundary CFT. Some
effort in this direction has been made in \cite{BPT}.

\paragraph{Acknowledgments:} The author would like to thank Benjamin Doyon and Andreas Fring for
useful discussions and comments on the  manuscript. She also
thanks Zoltan Bajnok for his interest in the work and, in
particular, for bringing reference \cite{Bajnok:2006} to her
attention. Finally, the author would like to thank the second
referee of the manuscript for his/her various constructive
comments and careful reading of the paper. This work has been
partially supported by EC Network EUCLID under contract number
HPRN-CT-2002-00325.

\appendix
\section{Formulae for the polynomials $R_{2n}^{e}$ and $R_{2n}^{o}$}
For the solutions already found in \cite{BPT} and listed in
(\ref{e1e4}), it is easy to see that
$R_{2}^{e}=R_{2}^{o}=R_{4}^{e}=R_4^{o}=0$. However, for higher
particle form factors non-trivial $R$-functions are found, which
become more involved as the particle number increases. Expressions
become simpler when written in terms of the polynomials:
\begin{equation}
\alpha_{i}=\sum_{k=0}^{[i/2]}(k+1) 4^{k}\sigma_{i-2k},\qquad
\beta_{i}=\frac{1}{2}\sum_{k=0}^{[i/2]}(k+1)(k+2)
4^{k}\sigma_{i-2k}
\end{equation}
We have constructed the following solutions\footnote{Notice that
the formula for $R_{16}^{e}$ given in the next page can only be
found in the ArXiv version of the manuscript. For shortness it has
been removed in J. Phys. A. published version of the manuscript.
This is however the only difference between the two documents.}:
\begin{eqnarray}
R_6^{e} &=& -1,\nonumber\label{r11}\\
R_8^{e}&=& \alpha_2^2,\nonumber\\
R_{10}^{e}&=& ( 8 ( 2 + {{\alpha }_2})  - {{\alpha }_4} )
{{\alpha}_6} +
  \beta_2
 ( {\alpha }_4^2 - 2{{\alpha }_8} )  + {{\alpha
 }_{10}},\\
R_{12}^{e}&=& -\beta_2{{\alpha }_6}(\beta_4{\alpha }_6-(( 16 +
{{\alpha }_2} ) {{\alpha }_8}   - 3{{\alpha }_{10}})
  )+ {\alpha }_6^3 - \beta_2^2 ( 8(2{{\beta }_2} + {{\alpha }_4} ) {{\alpha }_8} -
     2\beta_4 {{\alpha }_{10}}+ \beta_2
     \alpha_{12}),\nonumber \\
 R_{14}^{e}&=&{\left( 8 + {{\alpha }_2} \right)
}^2{\alpha }_8^3 +
  {\alpha }_8^2\left( {\alpha }_6^2 - {{\alpha }_2}{{\alpha }_6}{{\beta }_4} - 4{{\beta }_2}{\beta }_4^2
  \right)+2{\alpha }_6^2\left( 4{{\alpha }_{10}} - {{\alpha }_{12}} \right) {{\beta }_4}\nonumber\\
  && -8{{\alpha }_{10}}{\beta }_4^2\left( \left( 2 + {{\alpha }_2} \right) {{\alpha }_6} + 2{{\beta }_2}{{\beta
  }_4}\right)+ \left( 64\left( 28 + {{\alpha }_2}\left( 6 + {{\beta }_2} \right)  \right)  + {\alpha }_4^2 \right) {{\alpha }_8}
  {{\alpha }_{10}}{{\beta }_2}\nonumber\\
  && +(4 \left( 128 + {{\alpha }_2}\left( 48 + 5{{\alpha }_2} \right)  \right) {{\alpha }_4} -
  2\left( 8 + {{\alpha }_2} \right)  -
  {{\alpha }_6}\left( 4{{\alpha }_2} + {{\beta }_4}
  \right)){{\alpha }_8}{{\alpha }_{10}}\nonumber\\
  && +\left( 1792 + 4{{\alpha }_2}\left( 16\left( 6 + {{\beta}_2} \right)  + {{\alpha }_4} \right)  - 12\alpha_{0}{{\alpha }_6} +
     {{\alpha }_8} \right) {\alpha }_{10}^2+ \left( 8{{\alpha }_{12}} - {{\alpha }_{14}} \right) {{\beta }_2}{\beta
     }_4^3 \nonumber\\
     &&  + \left( 2{{\alpha }_2}{{\alpha }_{12}} + {{\alpha }_{14}} \right){{\alpha }_6}
   {\beta }_4^2+ {{\alpha }_{10}}{{\alpha }_{12}}\left(  \left( 4 -{{\alpha }_2} \right){{\alpha }_4}  + 3{{\alpha }_6} -
     4\left(  7{\alpha }_2^2 + 84{{\beta }_2}-48 \right)  \right) \nonumber\\
     && +
  \left( 8 + {{\alpha }_2} \right)
   \left( {{\alpha }_6} - \left(8 + 3{{\beta }_2} \right) {{\beta }_4}
   \right){{\alpha }_8}{{\alpha }_{12}}+ \left( 208 + 3{{\alpha }_2}\left( 16 + {{\alpha }_2} \right)\right) {\alpha }_{12}^2
   \nonumber\\&&+
  \left( 16\left( 16 + {{\alpha }_4} \right)  + {{\alpha }_2}\left( 64 + 3{{\beta }_4} \right)  -
     2{{\alpha }_6} \right) {{\alpha }_8}{{\alpha }_{14}}+
  \left( 8\left( 8 + {{\alpha }_2} \right)  - {{\alpha }_4} \right) {{\alpha }_{10}}{{\alpha }_{14}} \nonumber \\
     && -
  \left( 16 + 3{{\beta }_2} \right) {{\alpha }_{12}}{{\alpha }_{14}} + {\alpha
  }_{14}^2,
  \end{eqnarray}
  \begin{eqnarray}
  R_{16}^{e} &=&(({{\alpha }_2}{{\alpha }_4} - {{\alpha }_6} -16\left( 4 + 3{{\beta }_2} \right)) {{\alpha }_6} -
  {\left( 8 + {{\alpha }_2} \right) }^2){\alpha }_8^2\alpha_{10}^2 +
  \left( 16\left( 7 + {{\alpha }_2} \right)  + {{\alpha }_4} \right) {{\alpha }_6}{\alpha }_{10}^3 \nonumber\\&&+
  2\left( 8 + {{\alpha }_2} \right) {{\alpha }_8}{\alpha }_{10}^3 - {\alpha }_{10}^4 -
  2^{8}\left(64 + {{\beta }_2}\left( 16 + {{\beta }_2}\left( 20 + 17{{\beta }_2} \right) \right)  \right) {{\alpha }_4}
   {{\alpha }_{10}}{{\alpha }_{12}} \nonumber\\&&
   - 128(\left(-8 + {{\beta }_2}\,\left( 2 + 7\,{{\beta }_2} \right)\right) {\alpha }_4^2
  + 2\left(16 + {{\beta }_2}\,\left( -4 + {{\beta }_2}\,\left( -3 + 2\,{{\beta }_2} \right)  \right)\right)
{{\alpha}_6}){{\alpha }_{10}} {{\alpha }_{12}} \nonumber\\&&
   - 4(16\left( -8 + {{\alpha }_2}\left( 14 + 3{{\alpha }_2} \right)  \right) + \left( 4 + 5{{\alpha }_2} \right) {\alpha }_4){\alpha }_4{{\alpha }_6}{{\alpha }_{10}}
   {{\alpha }_{12}}+
  {\alpha }_6^3{{\alpha }_{10}}{{\alpha }_{12}}    \nonumber\\&&
  - (4\left( -160 + \left( -24 + {{\alpha }_2} \right) {{\alpha }_2} \right) + \left( -16 + {{\alpha }_2} \right) {{\alpha
}_4}){\alpha}_6^2{{\alpha }_{10}}{{\alpha }_{12}} \nonumber\\&&+
  \left( 8 + {{\alpha }_2} \right) \left( 8\left( 16\left( 8 + {{\alpha }_4} \right)   +
        {{\alpha }_2}\left( 48 + 2{{\alpha }_2} + {{\alpha }_4} \right)  \right) + \left( 24 + {{\alpha }_2} \right) {{\alpha }_6}
     \right) {{\alpha }_8}{{\alpha }_{10}}{{\alpha }_{12}} \nonumber\\&& +
  (16\left(80 + {{\beta }_2}\,\left( 32 + {{\beta }_2}\,\left( 11 + {{\beta }_2} \right)  \right) \right)
  + 2{{\alpha }_2}\left( 12 + {{\alpha }_2} \right) {{\alpha }_4} -
  \left( 40 + 3{{\alpha }_2} \right) {{\alpha }_6}){\alpha }_{10}^2{{\alpha }_{12}} \nonumber\\&&
  +4{{\alpha }_6}\left( 96\left( 80 + {{\alpha }_4} \right)  +
     {{\alpha }_2}\left( 2368 - {{\alpha }_2}\left( -192 + {{\alpha }_4} \right)  + 12{{\alpha }_4} + {{\alpha }_6} \right)  \right)
     {\alpha }_{12}^2\nonumber\\&& + 12{\left( 8 + {{\alpha }_2} \right)
}^2\left( 12 + {{\alpha }_2} \right) {{\alpha }_8}
   {\alpha }_{12}^2 - \left( 8 + {{\alpha }_2} \right) \left( 12 + {{\alpha }_2} \right) \left( 24 + {{\alpha }_2} \right)
   {{\alpha }_{10}}{\alpha }_{12}^2 \nonumber\\&&
   -2(2^{8}\beta_2
   \left( -80 + \left( -8 + {{\alpha }_2} \right) {{\alpha }_2} \right) {{\alpha }_4}
   - {\alpha }_6^2\left( {{\alpha }_6} + 4\,{\left( 4 + {{\beta }_2} \right) }^2 - \left( -8 + {{\beta }_2} \right) \,{{\beta }_4} \right))
   {{\alpha }_8}{{\alpha }_{14}}\nonumber\\
&&
      +
 \left( 32\left( 64 + {{\beta }_2}\left( 2 + {{\beta }_2} \right) \left( 8 + 5{{\alpha }_2} \right)  \right)
         + {{\alpha }_4}\left( -2^{9} + 44{{{\alpha }_2}}^2 + 3{{\alpha }_2}\left( 64 + {{\alpha }_4} \right)  \right) \right) {{\alpha }_6} {{\alpha }_{10}}{{\alpha }_{14}} \nonumber\\&&
   -
     \left( 28\left( 8 + {{\alpha }_2} \right)  + 3\beta_4 \right) {{\alpha }_6^{2}} {{\alpha }_{10}}{{\alpha }_{14}}   -
  8\left( 8 + {{\alpha }_2} \right) \left( 80 + {{\alpha }_2}\left( 22 + {{\alpha }_2} \right)  \right) {{\alpha }_8}
   {{\alpha }_{10}}{{\alpha }_{14}}\nonumber\\&&
   - \left( 8 + {{\alpha }_2} \right) (\left( 24 + {{\alpha }_2} \right) {{\alpha }_4}
  + 4{{\alpha }_6}){{\alpha }_8}
   {{\alpha }_{10}}{{\alpha }_{14}}-
  16\left({{\beta }_2}\left( 8 + 3{{\beta }_2}\left( 4 + {{\beta }_2} \right)  \right) -32\right) {{\alpha }_4}
   {\alpha }_{14}^2 \nonumber\\&& + \left(
     4\left( -32 + 2{{\alpha }_4} + {{\alpha }_6} \right)  -{{\alpha }_2}\left( 4\left( 8 + {{\alpha }_2} \right)
     + {{\beta }_4} \right) \right) {\alpha }_{10}^2{{\alpha }_{14}}\nonumber\\&& +
  4\left( 32\left( 2^{9} + {{\beta }_2}\left( 4 + {{\beta }_2} \right) \left( 88 + {{\beta }_2}\left( 20 + 9{{\beta }_2} \right)  \right) \right)
     + \beta_2 \left( 40 + 7{{\alpha }_2} \right) {\alpha }_4^2
 \right)
   {{\alpha }_{12}}{{\alpha }_{14}}\nonumber\\
  &&+
     16{{\alpha }_4}\left( -320 + 3{{\beta }_2}\left( -16 + {{\beta }_2}\left( 20 + 7{{\beta }_2} \right)  \right)  \right){{\alpha }_{12}}{{\alpha
     }_{14}}\nonumber\\&&
    - (4\left( 896 + {{\beta }_2}\left( 528 + {{\beta }_2}\,\left( 112 + {{\beta }_2} \right)  \right) \right)
    - \left( -192 + \left( -32 + {{\alpha }_2} \right) {{\alpha }_2} \right) {{\alpha }_4}){{\alpha }_6}
   {{\alpha }_{12}}{{\alpha }_{14}}\nonumber\\&& - ((\left( {{\alpha }_2}-8 \right) {\alpha }_6^2
   +
  3{\left( 8 + {{\alpha }_2} \right) }^2\left( 16 + {{\alpha }_2} \right) {{\alpha }_8})
  -
  \left( 8 + {{\alpha }_2} \right) \left( 72 + 5{{\alpha }_2} \right) {{\alpha }_{10}}){{\alpha }_{12}}{{\alpha }_{14}}\nonumber\\&& -
  16\left( 2^{10} + {{\beta }_2}\,\left( 896 + {{\beta }_2}\,\left( 416 + {{\beta }_2}\,\left( 108 + 13\,{{\beta }_2} \right)  \right)  \right) \right)
  {\alpha }_{14}^2 \nonumber\\&&+ \left( 2432 + 24{{\alpha }_4} + {{\alpha }_2}\left( 688 + 52{{\alpha }_2} + 5{{\alpha }_4} \right)  -
     2{{\alpha }_6} \right) {{\alpha }_6}{\alpha }_{14}^2 +
  3{\left( 8 + {{\alpha }_2} \right) }^2{{\alpha }_8}{\alpha }_{14}^2\nonumber\\&& -
  4\left( 8 + {{\alpha }_2} \right) {{\alpha }_{10}}{\alpha }_{14}^2 +
  2{\left( 8 + {{\alpha }_2} \right) }^2\left( \left( 8 + {{\alpha }_2} \right) {{\alpha }_8} - {{\alpha }_{10}} \right)
   {{\alpha }_{10}}{{\alpha }_{16}}\nonumber\\&& + {\left( 8 + {{\alpha }_2} \right) }^2
   \left( -8\left( 2^{8} + {{\alpha }_2}\left( 80 + 8{{\alpha }_2} + {{\alpha }_4} \right)  \right)  +
     \left( 24 + {{\alpha }_2} \right) {{\alpha }_6} \right) {{\alpha }_{12}}{{\alpha }_{16}} \nonumber\\&&+
  {\left( 8 + {{\alpha }_2} \right)}^2\left( 1152 + 8{{\alpha }_4} +
     {{\alpha }_2}\left( 336 + 28{{\alpha }_2} + 3{{\alpha }_4} \right)  - 4{{\alpha }_6} \right) {{\alpha }_{14}}
   {{\alpha }_{16}} - {\left( 8 + {{\alpha }_2} \right) }^4{\alpha }_{16}^2
   \nonumber\\&&-
  2^{10}\left( 896 + {{\alpha }_2}\left( 496 + {{\alpha }_2}\left( 100 + 7{{\alpha }_2} \right)  \right)  \right) {{\alpha }_{10}}
   {{\alpha }_{12}}{{\beta }_2} -32{\alpha }_4^3{{\beta }_2}(2{{\alpha }_{10}}{{\alpha }_{12}}
   +{{\alpha }_8}{{\alpha }_{14}})\nonumber\\&& -
  \left( 16 + 3{{\alpha }_2} \right) {\alpha }_4^2{\alpha }_{14}^2{{\beta }_2} +
  2^{14}{{\alpha }_8}{{\alpha }_{14}}{\beta }_2^2\left( 2 + {{\beta }_2} \right)  -
  2^{8}{\alpha }_4^2{{\alpha }_8}{{\alpha }_{14}}\left( -8 + {{\alpha }_2}{{\beta }_2} \right) \nonumber\\&& +
  64\left( -2^{8}\left( 88 + 3{{\alpha }_4} \right)  -
     2{{\alpha }_2}\left( 7936 + 2{{\alpha }_2}\left( 1120 + {{\alpha }_2}\left( 144 + 7{{\alpha }_2} \right)  \right)   \right) \right){\alpha }_{12}^2  \nonumber\\&&+
  64\left(
        \left( 320 + {{\alpha }_2}\left( 72 + 5{{\alpha }_2} \right)  \right) {{\alpha }_4}  -\left( 6 + {{\alpha }_2} \right)
        {\alpha }_4^2{{\beta }_2}\right){\alpha }_{12}^2+12{\alpha }_4^2{{\beta }_2}{{\alpha }_{10}}{{\alpha }_{14}}{{\beta }_4}\nonumber\\&&
     +
  \left( -16\left( 16 + {{\alpha }_2}\left( 9 + {{\alpha }_2} \right)  \right) {{\alpha }_4} -
     {\alpha }_4^2{{\beta }_2} - 64\left( 16 + {{\beta }_2}\left( 8 + {{\beta }_2}\left( 4 + {{\beta }_2} \right)  \right)
        \right)  \right){\alpha }_{10}^3 \nonumber\\&&
         - 16\left( \left( 2 + {{\alpha }_2} \right) {{\alpha }_4} -8\left( 28 + 5{{\alpha }_2} \right)\right)
     {{\alpha }_6}{{\alpha }_8}{{\alpha }_{14}}{{\beta }_4} +
  4\left(\left({{\alpha }_4} -32\right) {{\beta }_2} -64\right) {{\beta }_4}{{\alpha }_8}{\alpha }_{10}^2\nonumber\\&& +
 4\left(
    4{{\alpha }_4}\left({{\beta }_2}\left( 12 + 11{{\beta }_2} \right) -32 \right)  +
    32\left( 64 + {{\beta }_2}\left( 32 + {{\beta }_2}\left( 18 + 5{{\beta }_2} \right)  \right)  \right)  \right)
    {{\alpha }_{10}}{{\alpha }_{14}}{{\beta }_4}\nonumber
        \\&&
    +
  8\left(  -4\left( 48 + {{\alpha }_4} \right)  +
        {{\alpha }_2}\left( -32 + 2{{\alpha }_2} + {{\alpha }_4} \right)  - {{\alpha }_6} \right) {{\alpha }_6^{2}}{{\alpha }_8}{{\alpha }_{12}} +
  2{\left( 8 + {{\alpha }_2} \right) }^2{\alpha }_8^2{{\alpha }_{14}}{{\beta }_6} \nonumber\\&&+
  64\left(
     \left( 2\left( -48 + {{\alpha }_4} \right)  + {{\alpha }_2}\left( -8 + 2{{\alpha }_2} + {{\alpha }_4} \right)  \right)
      \right){{\beta }_4}{{\alpha }_6}{{\alpha }_8}{{\alpha }_{12}}
      - 8{\left( 8 + {{\alpha }_2} \right) }^2{\alpha }_8^2{{\alpha }_{12}}{{\beta }_6}\nonumber\\&& -
  \left( 8 + {{\alpha }_2} \right) \left( 8\left( 48 + {{\alpha }_4} \right)  +
     {{\alpha }_2}\left( 16\left( 9 + {{\alpha }_2} \right)  + 3{{\alpha }_4} \right) - 4{{\alpha }_6} \right) {{\alpha }_{10}}
   {{\alpha }_{16}}{{\beta }_6} \nonumber\\&& - {\left( 8 + {{\alpha }_2} \right) }^2{{\alpha }_8}{{\alpha }_{16}}{\beta }_6^2 +
  \left( -{{\alpha }_6} + {{\beta }_2}{{\beta }_4} \right) {\beta }_6^2
   \left( 4{\alpha }_{10}^2 + \left( 16{{\alpha }_{12}} - 8{{\alpha }_{14}} + {{\alpha }_{16}} \right) {{\beta }_6}
   \right).
  \end{eqnarray}
  The polynomials $R_{2n}^{o}$ can be obtained from the polynomials
$R_{2n-2}^{e}$ by substituting $\alpha_k \rightarrow
\alpha_{k+1}$,  $\beta_k \rightarrow \beta_{k+1}$, multiplying by
a global factor $-1$ and introducing as many factors $\sigma_1$ as
needed in order to achieve that every term in $R_{2n}^{o}$ is a
product of exactly $n-3$ elementary symmetric polynomials. For
example:
\begin{eqnarray}
  R^{o}_{6} &=& 0, \nonumber\\
  R^{o}_{8} &=& \sigma_1,\nonumber\\
  R^{o}_{10} &=& -\alpha_3^2,\\
  R^{o}_{12} &=& -( 8 ( 2\alpha_{1} + {{\alpha }_3})  - {{\alpha }_5} )\alpha_{1}
{{\alpha}_7} -
  \beta_3
 ( {\alpha }_5^2 - 2\alpha_{1} {{\alpha }_9} )  + \alpha_{1}^{2} {{\alpha }_{11}},\nonumber\\
R_{14}^{o}&=& \beta_3{{\alpha }_7}(\beta_5{\alpha }_7-(( 16
\alpha_1+ {{\alpha }_3} ) {{\alpha }_9}   - 3\alpha_1 {{\alpha
}_{11}})
  )+ \alpha_1{\alpha }_7^3 + \beta_3^2 ( 8(2{{\beta }_3} + {{\alpha }_5} ) {{\alpha }_9} -
     2\beta_5 {{\alpha }_{11}}+ \beta_3
     \alpha_{13}),\nonumber
\end{eqnarray}
and so on. Notice that $\alpha_1=\sigma_1$.

\small

\end{document}